\documentclass[showpacs,twocolumn,amsmath,amssymb,longbibliography]{revtex4-1}

\usepackage{xifthen}% provides \isempty test

\newcommand{\refAppendix}[6]{#1
  \ifthenelse{\isempty{#2}}%
    {}% if #2 is empty
    {\protect\cite{#2}}% if #1 is not empty
    #3\protect\ref{#4}#5#6\xspace
}

\usepackage{forloop,ifthen} % for loops and if statements
\usepackage{multirow} % merge rows and columns
\usepackage{ esint }

\usepackage{amsmath} % need for subequations

    \usepackage{graphicx}
    \usepackage{transparent}
    \usepackage{fancybox}  
    \usepackage{color}
    \usepackage{dcolumn}
    \usepackage{bm}
    \usepackage{xspace}
    \usepackage[colorlinks=true,     linkcolor={red!70!black},
    citecolor={green!50!black}, urlcolor={blue!50!black}, pdfborder={0 0 0}]{hyperref}

\usepackage{calc}

    \newcommand{\VEC}[1]{\mbox{\boldmath${#1}$}}
    \newcommand{\CO}[1]{}%SuppressText
    \newcommand{\emphLabel}[1]{\textbf{({#1})}}

\usepackage[usenames,dvipsnames]{xcolor}
\usepackage{ amssymb }

\newcommand{\ps}{phase space\xspace}

\newcommand{\suppress}[1]{}%SuppressText

\newcommand{\emphCaption}[1]{{\bf{#1}}}

\begin{document}

\title{Isospectral mapping for quantum systems with energy point spectra \\to  polynomial quantum harmonic oscillators}

\author{Ole Steuernagel${}^{1}$ and Andrei B. Klimov${}^{2}$}

\affiliation{${}^{1}${School of Physics, Astronomy and Mathematics,~University of Hertfordshire, Hatfield, AL10 9AB, UK}\\
${}^{2}$Departamento de F\'isica, Universidad de Guadalajara, 44420 Guadalajara, Jalisco, Mexico}

\date{\today}

\begin{abstract}
  % Harmonic quantum oscillators are building blocks of quantum theory, particularly field
  % theory.  Here
  We show that a polynomial $\hat {\cal H}_{(N)}$ of degree~$N$ of a harmonic oscillator
  hamiltonian allows us to devise a fully solvable continuous quantum system for which the
  first $N$ discrete energy eigenvalues can be chosen at will. In general such a choice
  leads to a re-ordering of the associated energy eigen\-functions of $\hat {\cal H}$ such
  that the number of their nodes does not increase monotonically with increasing level
  number. Systems $\hat {\cal H}$ have certain `universal' features, we study their basic
  behaviours.
\end{abstract}

% \pacs{03.65.-w, %Quantum mechanics 
%  03.65.Ta %Foundations of quantum mechanics
% }

\maketitle

\section{Introduction \label{sec_intro}}

Unlike for finite discrete systems, for continuous quantum systems it is generally hard to work out
their energy spectrum, and given an energy spectrum, it is generally hard to write down a
hamiltonian $\hat {\cal H}$ of a continuous system with that spectrum.

It is therefore noteworthy that, given a finite arbitrary set of $N$ real values, a
continuous one-dimensional quantum system's hamiltonian $\hat {\cal H} (\hat x, \hat p)$,
with this set as its first $N$ energy eigenvalues, can be devised. This is shown here by
explicit construction of a formal hamiltonian using real
polynomials~${\cal P}(\hat h) = \hat {\cal H}_{(N)}$ of $N$-th degree of the harmonic
quantum oscillator hamiltonian~$\hat h (\hat x, \hat p)$. For polynomials of low degree
$N$ such hamiltonians~$\cal H$ can arise as effective descriptions of
fields~\cite{Bender_PRL08}, oscillating beams~\cite{Berry_JPAMG86},
nano-oscillators~\cite{Jacobs_PRL09},
Kerr-oscillators~\cite{Dykman_PRB05,Yurke_JLT06,Bezak_AP16,Zhang_PRA17,Oliva_Kerr_18}, and
cold gases~\cite{Greiner_NAT02}.

In this work we primarily consider conservative one-dimensional quantum mechanical bound
state systems of one particle with mass~$M$ subjected to a trapping potential~$V(x)$, i.e.,
hamiltonians of the form
\begin{eqnarray}
  \label{eq:_QM_hamiltonian}
\hat H = \frac{\hat p^2}{2 M} + V(\hat x) 
\end{eqnarray}
as reference systems.

In Section~\ref{sec_Map_2_Hosc} we introduce the mapping from $\hat H$ to $\hat {\cal H}$
and we stress that this mapping can reorder wave functions violating the Sturm-Liouville
rule for monotonic energy-level ordering of quantum mechanical systems.  We then show in
Section~\ref{sec_NumericalImplementation} that the mapping for an increasing number~$N$ of
energy levels of a fixed $\hat H$ to a family of mapped systems $\hat {\cal H}_{(N)}$,
which share these energy level values, does not generally converge in the limit of
large $N$. In Section~\ref{sec_continuous_deformation} we investigate how continuous
deformations of the potential~$V(x)$ in Eq.~(\ref{eq:_QM_hamiltonian}) affect the formal
hamiltonian~${\cal H}$. We consider deformations which take systems that do not exhibit
tunnelling behaviour to ones that do and we consider the transition from one- to
multi-well systems. We finally comment on the phase space behaviour of~$\hat {\cal H}$
and the reshaping of states when mapping between $\hat H$ and~$\hat {\cal H}$ in
Sections~\ref{sec_Phase_Space_J} and~\ref{sec_Mapped_States}, before we conclude.

\section{Mapping to polynomials of harmonic oscillator hamiltonians \label{sec_Map_2_Hosc}}

The (dimensionless) harmonic oscillator hamiltonian is given by
$ \hat h = \frac{\hat p^2 }{2} + \frac{\hat x^2 }{2} , $ where we set Planck's reduced
constant $\hbar$, the spring constant and mass~$M$ of the oscillator all equal to
`1'. Expressing position~$\hat x = \frac{1}{\sqrt{2}}(\hat b^\dagger + \hat b)$ and
momentum~$\hat p = \frac{i}{\sqrt{2}}(\hat b^\dagger - \hat b)$ in terms of bosonic
creation operators~ $\hat b^\dagger$ and annihilation operators $\hat b$, that fulfil the
commutation relation $[ \hat b, \hat b^\dagger ] = \hat 1$ and form the number operator
$ \hat b^\dagger \hat b = \hat n$, allows us to write
$ \hat h = \hat b^\dagger \hat b + \frac{\hat 1 }{2} = \hat n + \frac{\hat 1 }{2} $.

Using~$\hat h$ as the argument of a polynomial~${\cal P}_{(N)}$ of order $N$ with real
coefficients~$a_j$ yields the formal hamiltonian
\begin{eqnarray}
  \label{eq:_Polynom}
\hat {\cal H}_{(N)} \equiv {\cal P}_{(N)} (\hat h) = \sum_{j=1}^N a_j \hat h^j .
\end{eqnarray} 

Its energy spectrum derives from the mapping of the harmonic oscillator spectrum
$h_n = n + \frac{1}{2}$ to
\begin{eqnarray}
  \label{eq:_H_spectrum}
E_n \equiv \langle \phi_n | \hat {\cal H} | \phi_n \rangle = {\cal P}(h_n) .
\end{eqnarray}

Here the eigenvalues~$n=0,1,2,...,\infty$ of the number operator~$\hat n$ label the
harmonic oscillator eigenfunctions
\begin{eqnarray}
  \label{eq:_eigenstates}
  \phi_n(x) =  \langle x | \frac{(b^\dagger)^n}{\sqrt{n!}} | 0 \rangle = \frac{1}{\sqrt{2^n n!\sqrt{\pi}}} e^{-\frac{x^2}{2}} \eta_n(x), \qquad
\end{eqnarray}
where $\eta_n$ are the Hermite polynomials.

We note that construction~(\ref{eq:_Polynom}) renders ${\cal H}$ fully solvable since all
its wave functions (and also associated phase space Wigner distribution functions) are
inherited from the harmonic oscillator.

Moreover, the associated classical hamiltonians ${\cal H}(x,p)={\cal P}(\frac{r^2}{2})$
are functions of $r=\sqrt{x^2+p^2}$ alone; their quantum
versions,~$\hat{\cal H} = {\cal H}(\hat x, \hat p)$, inherit this phase space symmetry and
obey probability conservation on the energy contours of the system (concentric circles
around the origin, i.e. invariance under O(2) rotations)~\cite{Oliva_Kerr_18}.

Such probability conservation on energy contours is not generic, it is a special case for
systems of the form~$\hat {\cal H}$, see Section~\ref{sec_Phase_Space_J}.

\subsection{Reordering of energy eigenfunctions \label{subsec_reorder}}

The Sturm-Liouville rule for quantum mechanical hamiltonians~(\ref{eq:_QM_hamiltonian}) states that
the ground state~$\phi_0$ is node-free and excited states~$\phi_n$ have $n$ nodes. It is based on
the Sturm-Liouville theorem which applies to second order differential equations but not here, since
$\hat {\cal H} = {\cal P}(\hat h)$ can be of high order in $\hat p$. Consequently we find that the
value of $E_n$ can be equal to or greater than $E_{n+1}$.

Below we show that the first $N$ energy eigenvalues
\{$ {\cal P}_{(N)}(h_n), n=0,...,N-1$\} can be constructed to coincide with any (random)
sequence of real numbers.

In 1986, Berry and Mondragon noticed energy-level degeneracies can occur for 1D-systems
which are quartic in momentum~\cite{Berry_JPAMG86}. That is a special case of our more
general observation about random level reordering.

More recently, the fact that the Sturm-Liouville ordering rule for wave function nodes can
be violated has been reported for the quasi-energy levels (in rotating-wave approx\-imation)
of driven Kerr systems~\cite{Dykman_PRB05,Zhang_PRA17}.

\begin{figure}[t]
  \hspace{-0.2cm}
  \begin{minipage}[b]{1.02\columnwidth}
%  "b" to have captions on the same line
    \includegraphics[width=0.319\columnwidth]{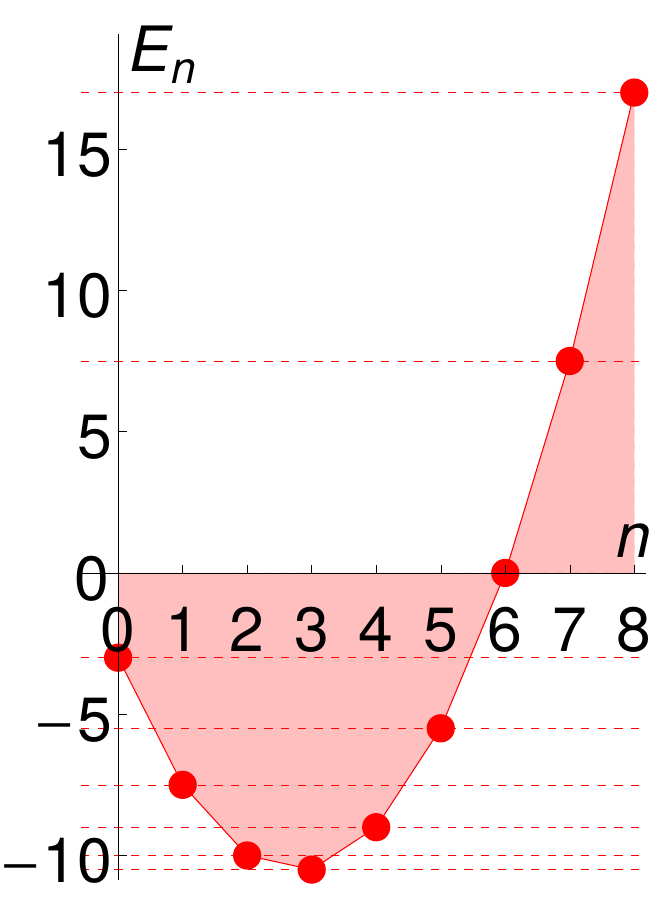}
    \put(-80,105){\rotatebox{0}{\emphLabel{a}}}
  %\!\!\!\!
  \includegraphics[width=0.663\columnwidth]{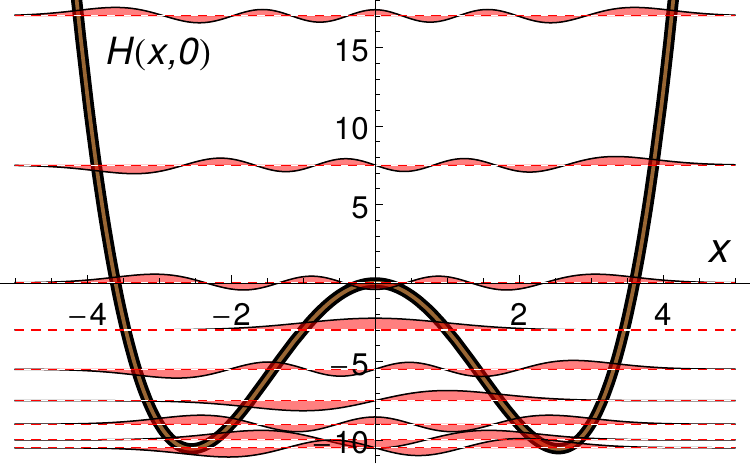}
    \put(-163,105){\rotatebox{0}{\emphLabel{b}}}
  \end{minipage}
  \caption{\emphCaption{Spectrum and energy eigenstates.}~\emphLabel{a}, the first nine
    energy eigenvalues of the harmonic oscillator $h_n = n+\frac{1}{2}$ for
    $n=0,1,2,...,8$ are mapped via the polynomial
    ${\cal P}(\xi) = \xi^2 - \frac{13}{2} \xi$ to $E_n={\cal P}(h_n)$. This map changes the
    order of the values $E_n$ such that the state with index $n=3$ has the lowest
    eigenenergy $E_3 = -\frac{21}{2}$.~\emphLabel{b}, the wave functions, with their
    ordinate off-set by $E_n$, are
    displayed together with the cross-section~${\cal P}(\frac{1}{2}x^2) = {\cal H}( x, 0)$ of the
    hamiltonian.
    \label{fig:Distributions}}
\end{figure}

For $\hat {\cal H} =  \hat h^2 - \frac{13}{2} \hat h$, this ordering violation is
demonstrated graphically in Fig.~\ref{fig:Distributions}; although this case is discussed
in Ref.~\cite{Bezak_AP16} the reordering is not mentioned there.

\subsection{Dialling up the spectrum \label{sec_dial_spectrum}}

We now show that we can dial up an arbitrary real point spectrum for the $N$ first energy
eigenstates of~$\hat {\cal H}$. Note, by `first' we mean the entries of the column-vector
$\VEC{E}_{(N)} = [E_0,E_1,...,E_{N-1}]^\top$ of Eq.~(\ref{eq:_H_spectrum}). Because of the
reordering, these are in general not the lowest lying energy values.

Rewriting Eq.~(\ref{eq:_Polynom}) in a suitable matrix form is achieved by casting the
energy values of $\hat h^j$ into the form of a square $N \times N$ `energy-matrix'
$[\VEC{\epsilon}_{(N)}]_{n,j} = (h_n)^j$.  The coefficient column-vector
$\VEC{a}_{(N)} = [a_1,a_2,...,a_N]^\top$ then, according to Eq.~(\ref{eq:_Polynom}), obeys
$\VEC{E}_{(N)} = \VEC{\epsilon}_{(N)} \cdot \VEC{a}_{(N)}$ where the dot stands for matrix
multiplication.

For instance, $\VEC{\epsilon}_{(5)}$ has the form
\begin{eqnarray}
  \label{eq:_EnergyMatrix}
  \VEC{\epsilon}_{(5)} = 
\begin{bmatrix}
 \frac{1}{2} & \frac{1}{4} & \frac{1}{8} & \frac{1}{16} & \frac{1}{32} \\[4pt]
 \frac{3}{2} & \frac{9}{4} & \frac{27}{8} & \frac{81}{16} & \frac{243}{32} \\[4pt]
 \frac{5}{2} & \frac{25}{4} & \frac{125}{8} & \frac{625}{16} & \frac{3125}{32} \\[4pt]
 \frac{7}{2} & \frac{49}{4} & \frac{343}{8} & \frac{2401}{16} & \frac{16807}{32} \\[4pt]
 \frac{9}{2} & \frac{81}{4} & \frac{729}{8} & \frac{6561}{16} & \frac{59049}{32} \\
\end{bmatrix}.
\end{eqnarray}
The determinant $|\VEC{\epsilon}_{(N)}|= \prod_{g=1}^{N-1}[g! (2 g + 1)] / 2^N$ is
non-zero, hence, $\VEC{\epsilon}_{(N)}$ can always be inverted: \emph{any} column vector
$\VEC{E}_{(N)}$ uniquely speci\-fies a coefficient vector~$\VEC{a}_{(N)}$ where

\begin{widetext}
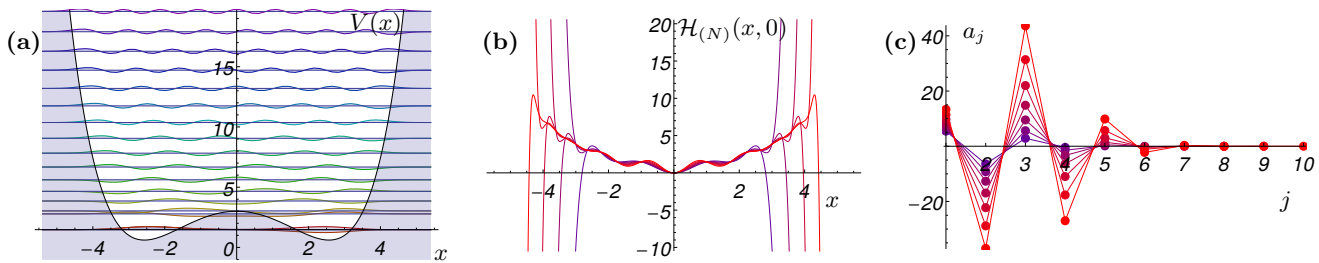
\begin{figure}[b]
  \hspace{-0.2cm}
  \begin{minipage}[b]{1.97\columnwidth}
%  "b" to have captions on the same line
    \includegraphics[width=\columnwidth]{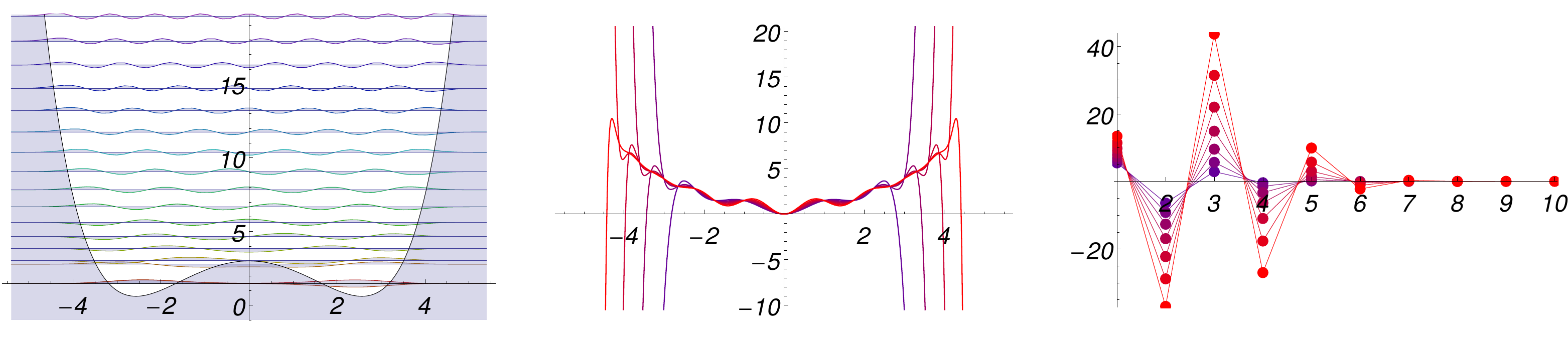}
    \put(-495,85){\rotatebox{0}{\emphLabel{a}}}
    \put(-315,85){\rotatebox{0}{\emphLabel{b}}}
    \put(-163,85){\rotatebox{0}{\emphLabel{c}}}
    %\put(-405,95){\rotatebox{0}{$V(x)$}}
    \put(-365,92){\rotatebox{0}{$V(x)$}}
    \put(-333,5){\rotatebox{0}{$x$}}
    \put(-241,90){\rotatebox{0}{${\cal H}_{(N)}(x,0)$}}
    \put(-185,25){\rotatebox{0}{$x$}}
    \put(-133,89){\rotatebox{0}{{$a_j$}}}
    \put(-13,25){\rotatebox{0}{$j$}}
    \caption{\emphCaption{Quantum mechanical potential~$V(x)$, its
        analogue~$\hat {\cal H}$ for varying orders and the associated expansion
        coefficients.}~\emphLabel{a}, for the %quantum mechanical
      hamiltonian~$\hat H$ its potential $V(x)=3-(2 x^2)/3+x^4/27+x^6/729$ is shown with
      its energy levels.~\emphLabel{b},~$\hat H$ is mapped to equivalent hamiltonian's
      $\hat {\cal H}_{(N)}$ of progressively increasing order ranging from $N=4$ to
      10. The increase in order is reflected in the colour coding (blue for $N$=4 to red
      for $N$=10) of the associated plots of the
      cross-sections~${\cal P}_{(N)}(\frac{1}{2}x^2) ={\cal H}_{(N)}( x, 0)$. We notice
      that the expansion coefficients ${\VEC a}_{(N)}$ do not settle, instead, across the
      board, they increase in magnitude with increasing $N$: see~\emphLabel{c} [employing
      the same color coding as~\emphLabel{b}]. Moreover, the expansion coefficients~$a_j$
      alternate in sign, this implies that for the potential~$V(x)$, displayed
        in~\emphLabel{b}, ${\cal H}_{(N)}(r)$ is negative (open downward) for large
      values of $r=\sqrt{x^2+p^2}$ if $N$ is even, and positive (closed) for odd~$N$.
      \label{fig:OscillatingCoefficients}}
\end{minipage}
\end{figure}
\end{widetext}

\begin{eqnarray}
  \label{eq:_a_from_E}
\VEC{a}_{(N)} = \VEC{\epsilon}^{-1}_{(N)} \cdot \VEC{E}_{(N)} \; 
\end{eqnarray}
and thus $\VEC{a}_{(N)}$ specifies a formal hamiltonian $\hat {\cal H}_{(N)}$ for which
the first $N$ eigenfunctions~$| \phi_n \rangle$ have energies $\VEC{E}_{(N)}$.

As mentioned before, this observation implies that~$\hat {\cal H}$ can be formed such that
any level is randomly assigned any real energy value:
$\langle \phi_n | \hat {\cal H} | \phi_n \rangle = E_n$.

However, in quantum mechanical systems we expect the Sturm-Liouville level ordering rule to be
obeyed.

In Subsection~\ref{sec_even_odd} we will find that in our construction
violations of the Sturm-Liouville level ordering rule arise spontaneously but, fortunately,
when we restrict the order of ${\cal P}_{(N)}$ to either even or odd values of $N$
(depending on the system~$\hat H$ considered) this irritating ordering violation is
absent.

In other words: if for a given potential~$V(x)$ the mapping of $H$ onto ${\cal H}_{(N)}$ does not
reproduce the lowest $N$ values of $\VEC{E}_{(N)}$, then a mapping onto ${\cal H}_{(N+1)}$ will.

Despite the fact that $H$ and the formal hamiltonian ${\cal H}_{(N)}$ share parts of their
energy spectrum, $V(x)$ and ${\cal H}_{(N)}(x,0)$ do not have any obvious functional
relationship, see Fig.~\ref{fig:OscillatingCoefficients}. This observation is reinforced
by the fact that the formal hamiltonian ${\cal H}_{(N)}$ is invariant under parity
transformations whereas $\hat H$ in general is not.

One is also free to generalise our approach, for example, by assigning energy values to
only some eigenfunctions~$\phi_n$. To this end one can strip out the $m$-th entry in
$\VEC{E}$ together with the $m$-th row in $\VEC{\epsilon}$ thus removing an assignment for
an `unwanted' state~$\phi_m$ [whose value would still be assigned implicitly through
Eq.~(\ref{eq:_H_spectrum})].  One then also has to strip out one column of
$\VEC{\epsilon}$ (together with the associated entry in $\VEC a$) to keep $\VEC{\epsilon}$
invertible. This column could, e.g., be the last ($N$-th) column in which case the
order of polynomial~$\cal P$ would be reduced by one.

\subsection{Shifting the ground state energy $E_0$ \label{subsec_shift_spectrum}}
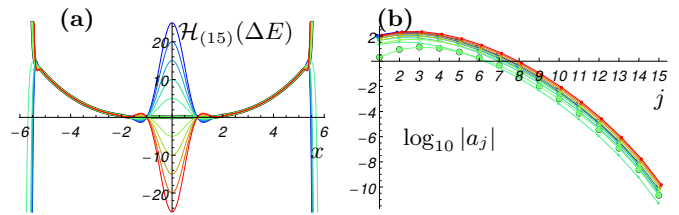
\begin{figure}[h]
  \hspace{-0.2cm}
  \begin{minipage}[b]{1.01\columnwidth}
    % "b" to have captions on the same line
    \includegraphics[width=\columnwidth]{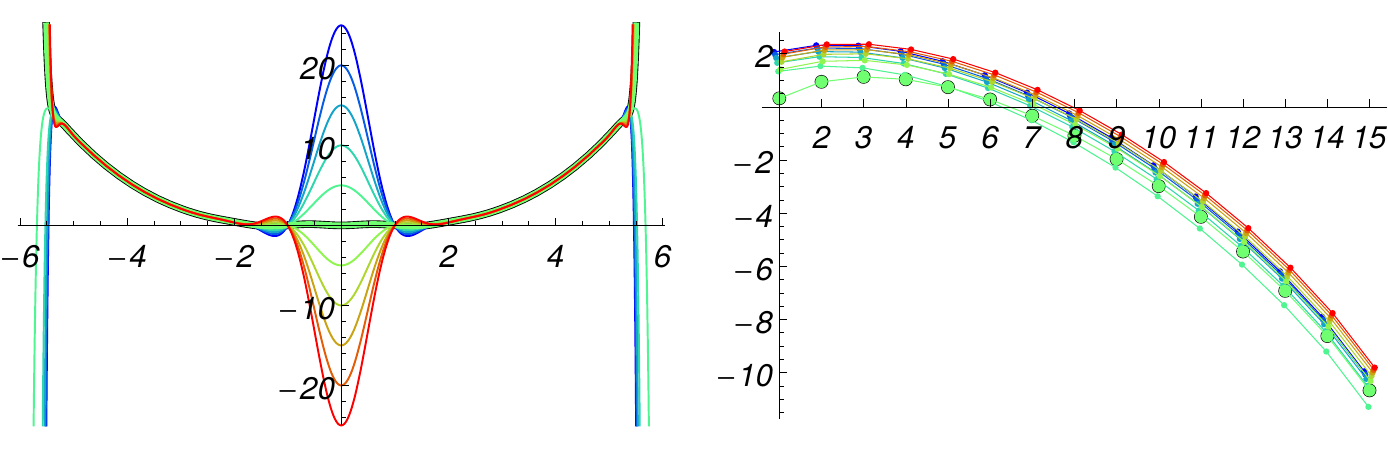}
    \put(-230,75){\rotatebox{0}{\emphLabel{a}}}
    \put(-110,75){\rotatebox{0}{\emphLabel{b}}}
    % \put(-235,65){\rotatebox{0}{\emphLabel{a}}}
    % \put(-90,65){\rotatebox{0}{\emphLabel{b}}}
    \put(-185,70){\rotatebox{0}{${\cal H}_{(15)}(\Delta E)$}}
    \put(-135,25){\rotatebox{0}{$x$}}
    \put(-100,30){\rotatebox{0}{$\log_{10}|a_j|$}}
    \put(-5,45){\rotatebox{0}{$j$}}
  \end{minipage}
  \caption{\emphCaption{Expansion coefficients vary with shift of energy.} The ground
    state energy $E_0$ for $\hat H$ with $V(x)=1-(2 x^2)/9+x^4/81+x^6/729$ is shifted
    (together with the rest of the energy spectrum) from -25 (red color) via 0 (green
    color, which is highlighted) to +25 (blue color). The associa\-ted curves for
    ${\cal H}(x,0)$ are shown in panel~\emphLabel{a} and the magnitude of the expansion
    coefficients $\log_{10}|a_j|, j=1,...,15$ in panel~\emphLabel{b}. One can see that the
    expansion coefficients tend to be small in magnitude for small energy offsets, in other
    words, setting $E_0=0$ (green lines) is a reasonable choice.
    \label{fig:Coefficients}}
\end{figure}

The expansion coefficients $\VEC a$ depend on the value of $E_0$.  We cannot shift the
harmonic oscillator's spectrum such that its ground state energy~$h_0 = 0$ since that
would render $|\VEC{\epsilon}|=0$ making $\VEC{\epsilon}^{-1}$ in Eq.~(\ref{eq:_a_from_E})
ill-defined. We will therefore from now on, for definiteness, set $E_0 = 0$: for further
justification see Fig.~\ref{fig:Coefficients}.

\section{Computational implementation of $\hat {\cal H}$ % = {\cal P}(\hat h)$
  and %some
  stability considerations\label{sec_NumericalImplementation}}

\subsection{Using exact fractions\label{sec_exact_fractions}}

Since we could not determine the general explicit form of $\VEC{\epsilon}^{-1}$ in
Eq.~(\ref{eq:_a_from_E}) we let a program determine it, in terms of exact fractions, to
avoid numeri\-cal instabilities associated with approximations of
$\VEC{\epsilon}^{-1}$. With modern computer algebra systems it is feasible to do so, in
seconds, for $N$-values of order $10^3$.

For the sake of numerical stability we found that the energy eigenvalues of $\hat H$,
which are used as input values, also have to be formally written in analytical form,
namely as fractions (e.g., $E_5 = 10.453$ should be written as $E_5 = 10453/1000$). Then,
even for fairly large values ${\cal O}(N) \approx 10^3$ can the analogue hamiltonian
$\hat {\cal H}_{(N)}$ be constructed safely using Eq.~(\ref{eq:_a_from_E}).  The
associated high order polynomial function ${\cal P}(x,p)$ is of order $2N$ in $x$ and $p$
and therefore becomes numerically unmanageable for moderate values
${\cal O}(N) \approx 10^2$, fortunately that does not affect the stability of the
underlying scheme encapsulated by Eq.~(\ref{eq:_a_from_E}).

\subsection{The coefficients do not settle down \label{sec_Unsettled}}

The expansion coefficients ${\VEC a}_{(N)}$ for a chosen quantum mechanical
hamiltonian~$H$ do not settle down with increasing order $N$ of the number of mapped
energy values ${\VEC E}_{(N)}$, see Fig.~\ref{fig:OscillatingCoefficients}~\emphLabel{c}; they alternate and increase in value with $N$. The underlying reason for this
behaviour is the fact that, with every added order $j$ of $\hat h^j$, momentum terms of
order $\hat p^{2j}$ are added. Their presence leads to so much `kinetic energy' added with
every order that even infinite-box potentials display oscillations of the
coefficients~$a_j$.

Therefore, typically, $\lim_{N\rightarrow \infty}{\cal H}_{(N)}$ does not exist.

\section{Smooth deformations of the potential and its effect on ${\cal H}_{(N)}$ \label{sec_continuous_deformation}}

\subsection{Even versus odd number of levels \label{sec_even_odd}}

In the mapping of $\hat H$ to $\hat {\cal H}_{(N)}$ we observed that for the fixed
potential portrayed in Fig.~\ref{fig:OscillatingCoefficients}~\emphLabel{a} the expansion
order~$N$ should be odd to avoid a downward open potential that violates the Sturm-Liouville
level ordering rule, see Fig.~\ref{fig:OscillatingCoefficients}~\emphLabel{b}. In general, a
fixed potential $V(x)$ requires either even or odd expansion orders~$N$ to achieve
this. This observation of an even-odd-$N$ bias is generic, since the oscillations of the
coefficients~$a_j$, as seen in Fig.~\ref{fig:OscillatingCoefficients}~\emphLabel{c}, is
typical.

The observation of such an `even-odd bias' in the desirable orders of the expansion
for~${\cal H}_{(N)}$ raises the question whether one can use this bias to devise
a criterion for grouping potentials into separate, inequivalent classes.

\subsection{Deformation from single well to deep double well potential \label{subsec_Single_double}}

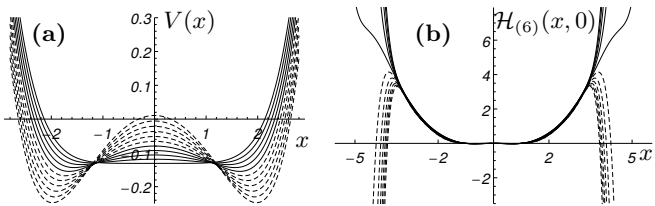
\begin{figure}[t]
  \hspace{-0.2cm}
  \begin{minipage}[b]{1.01\columnwidth}
%  "b" to have captions on the same line
    \includegraphics[width=\columnwidth]{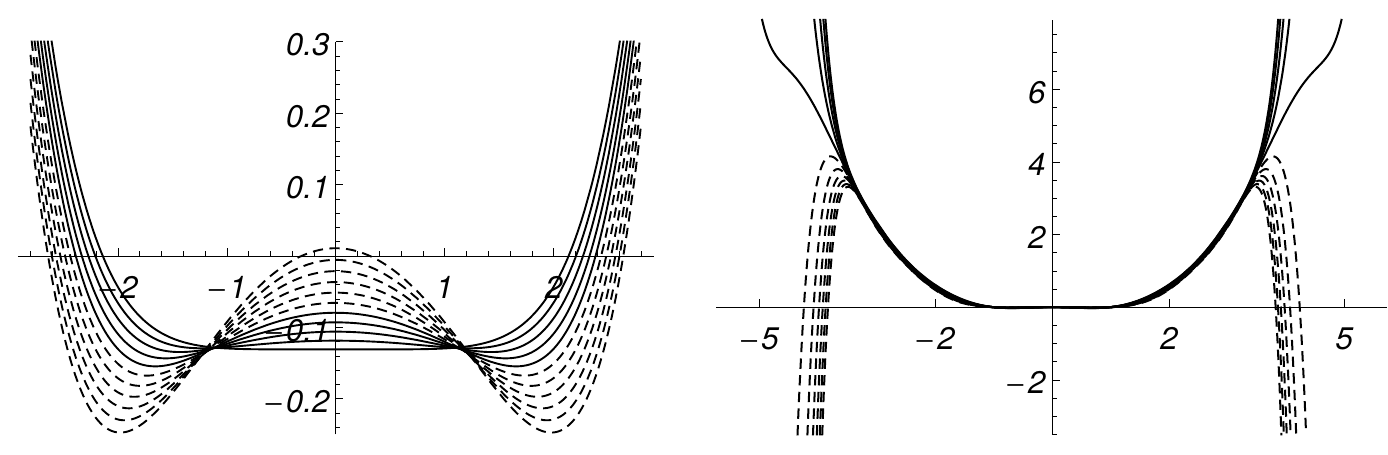}
    \put(-235,65){\rotatebox{0}{\emphLabel{a}}}
    \put(-90,65){\rotatebox{0}{\emphLabel{b}}}
    \put(-185,70){\rotatebox{0}{$V(x)$}}
    \put(-135,25){\rotatebox{0}{$x$}}
    \put(-60,70){\rotatebox{0}{${\cal H}_{(6)}(x,0)$}}
    \put(-5,20){\rotatebox{0}{$x$}}
  \end{minipage}
  \caption{\emphCaption{Expansion order bias switch in deformed potentials.}
    Panel~\emphLabel{a} displays potentials
    $V(x)=V_0\left[1-(2 x^2)/9+x^4/81\right] +x^6/729 - E_0$ where $V_0$ is incremented
    from 0 to 0.5 in ten equal steps of size $0.05$. This leads to a transition from
    single to double well potential where the groundstate (at shifted energy
    $\tilde E_0=0$) just starts to tunnel.  Panel~\emphLabel{b} displays the associated
    formal hamiltonians~${\cal H}_{(6)}$ and shows that an even to odd transition for the
    desirable expansion order~$N$ occurs.  For more details see main text.
    \label{fig:EvenVsOddCoefficients}}
\end{figure}

In the case of a continuous transition of single well to double well potentials, as
sketched in Fig.~\ref{fig:EvenVsOddCoefficients}, such an even-odd transition occurs once.
In this case we do not, for instance, witness a back-and-forth switching between even-odd
and odd-even biases with, say, every addition of the next higher eigenstate to the
tunnelling regime.

%\section{Spin systems\label{sec_Spin_systems}}
%\subsection{Even versus odd number of levels \label{sec_even_odd_SPIN}}
% I intend to study spin spectra in the form $E_m = m + L$ corresponding to ($\hat H = J_z$)
% versus $E_m = (m^2 - L^2)$ corresponding to ($\hat H = J_z^2$) and also coupled spins:
% $E_m = (m + \Delta(m) + L + \varepsilon), \Delta(m) = -\varepsilon $ when $\varepsilon$ is
% even and $\Delta(m) = +\varepsilon $ when $\varepsilon$ is odd, in all cases
% $m = -L, -L+1, ..., L$; in the last case $L$ even only.

\subsection{Multi-well systems\label{subsec_Multiwell_systems}}

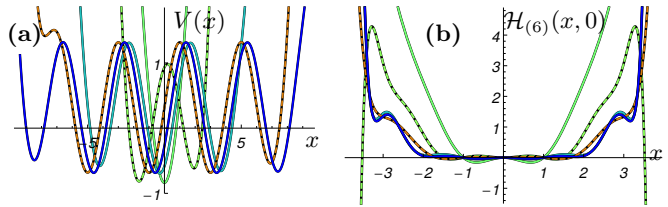
\begin{figure}[h]
  \hspace{-0.2cm}
  \begin{minipage}[b]{1.01\columnwidth}
%  "b" to have captions on the same line
    \includegraphics[width=\columnwidth]{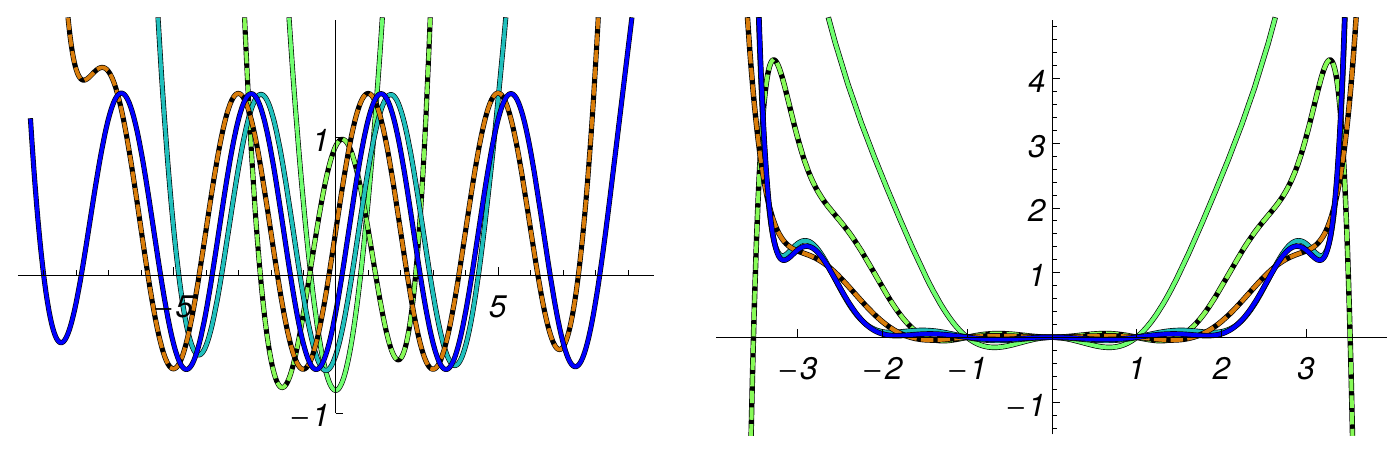}
    \put(-248,65){\rotatebox{0}{\emphLabel{a}}}
    \put(-90,65){\rotatebox{0}{\emphLabel{b}}}
    \put(-185,70){\rotatebox{0}{$V(x)$}}
    \put(-135,25){\rotatebox{0}{$x$}}
    \put(-60,70){\rotatebox{0}{${\cal H}_{(6)}(x,0)$}}
    \put(-5,20){\rotatebox{0}{$x$}}
  \end{minipage}
  \caption{\emphCaption{Expansion order bias in multi-well systems.}
    Panel~\emphLabel{a} displays potentials for which the number of adjacent wells
    increases from 1 to 5. Panel~\emphLabel{b} displays the associated formal
    hamiltonians~${\cal H}_{(6)}$, which are 
    depicted by the same coloring of the graph as in panel~\emphLabel{a}.
    This shows that a transition in the even-odd-$N$ bias for the
    desirable expansion order~$N$ occurs whenever another well is added:
    curves are dashed when ${\cal H}_{(6)}(x,0) < 0$ for large values of $x$.
    \label{fig:MultiWell}}
\end{figure}

Instead of deforming the potentials such that they form increasingly deeper wells, as
considered in Fig.~\ref{fig:EvenVsOddCoefficients}, we now consider systems with an
increase in the number of adjacent wells, see Fig.~\ref{fig:MultiWell}.

We map this to a sixth-order formal hamiltonian~$\hat {\cal H}_{(6)}$ and observe
even-odd-$N$ bias transitions whenever another well is added. 

Similarly to our finding reported in Fig.~\ref{fig:EvenVsOddCoefficients}, an increase in the
barrier height between adjacent wells does, however, not affect the observed even-odd-$N$
bias:

The even-odd-$N$ biases can be used to discriminate between different numbers of wells in
multi-well potentials but not between the strength of the barriers between the wells.

The question whether this can be a useful criterion to sensitively discriminate between
different types of hamiltonians in other contexts remains open.

\section{Phase space behaviour\label{sec_Phase_Space_J}}

% \section{Generically, probability is not conserved on energy contours in phase
%   space \label{subsec_NO_prob_conserv}}

The evolution of the Wigner's phase space distribution function in quantum phase space is
governed by the continuity equation
$\frac{\partial W}{\partial t} = - \VEC{\nabla} \cdot \VEC{J} $, where $\VEC J$ is
Wigner's current~\cite{Oliva_PhysA17}.

Hamiltonians of the form $\hat {\cal H} = {\cal P}(\hat h)$ have special dynamical
features: their \ps current follows circles concentric to the \ps origin; for a proof see
the Appendix of Ref.~\cite{Oliva_Kerr_18}. In other words, their \ps current is tangential
to the system's energy contours.

We now show that such a special alignment cannot be constructed for quantum mechanical systems
with anharmonic potentials~$V(x)$.

In reference~\cite{Oliva_PhysA17} it was shown that anharmonic systems exhibit
singularities of the associated phase space velocity field. This precludes the possibility
of mapping them to system whose dynamics can be described by the Poisson-bracket of
classical physics~\cite{Kakofengitis_PRA17}, but directional alignment of $\VEC J$ with the energy
contours is not ruled out.

For a contradiction, assume that the desired directional alignment is possible
for anharmonic quantum mechanical systems when using a correction to the current
field~$\VEC{\tilde J}$, where $\VEC{\nabla} \cdot \VEC{\tilde J} = 0$, to not affect the
dynamics.

Per assumption $\VEC{J} + \VEC{\tilde J}$ is aligned with the classical hamiltonian flow
in phase space and therefore shares its stagnation points. But at a stagnation point
$\partial_t W = 0$, yet we know that quantum and classical phase space current stagnation
points do not in general coincide~\cite{Kakofengitis_EPJP17}: anharmonic quantum
hamiltonians can therefore not feature phase space current fields aligned with their
energy contours, unless they are of form~$\hat {\cal H}$.

\section{Mapped states\label{sec_Mapped_States}}

In this work we implicitly considered the diagonalization of a wide variety of
hamiltonians~$H$ and their subsequent mapping to generic systems of the form~${\cal H}$.
The dia\-go\-nalization of a quantum hamiltonian is not a smooth transformation.

\begin{figure}[h]
  \hspace{-0.2cm}
  \begin{minipage}[b]{0.97\columnwidth}
%  "b" to have captions on the same line
    \includegraphics[width=\columnwidth]{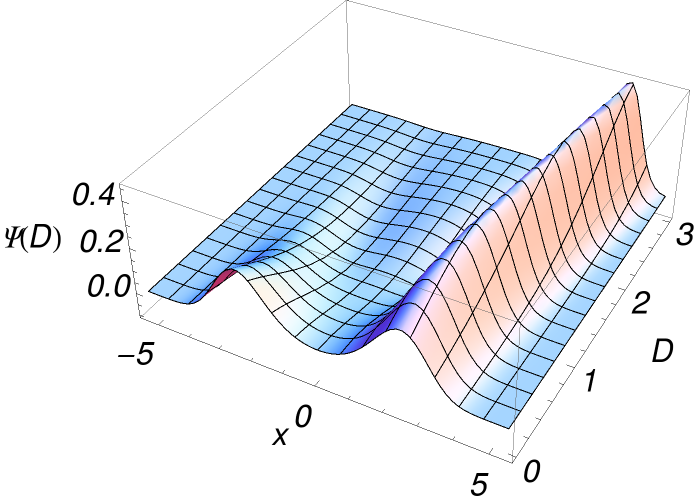}
    \caption{\emphCaption{Mapped `coherent states'.} Using Glauber coherent state coefficients
      for the energy eigenstates~$\psi_j(x)$ of
      potential $V(x)=3-(2 x^2)/3+x^4/27+x^6/729$, see Fig.~\ref{fig:OscillatingCoefficients}~\emphLabel{a}, 
      the mapped `coherent states' $\Psi(x,D) = \exp[-|D|^2/2]\frac{D^j}{\sqrt{j!}} \; \psi_j(x)$,
      displaced by~$D$, are
      formed. Their non-Gaussian shape illustrates that the mapping between $\hat H$ and $\hat {\cal H}$
      distorts features of wave functions.
      \label{fig:ConstipatedStates}}
\end{minipage}
\end{figure}

It is therefore of some interest to get a feeling for the distortions a state suffers when
mapping between hamiltonians~$H$ and~${\cal H}$. For illustration we consider the
distortions a gaussian Glauber-coherent state of system $\hat {\cal H}$ suffers, as a
function of displacement from the origin, when mapped to a double well system, see
Fig.~\ref{fig:ConstipatedStates}.

\section{Conclusions\label{sec_conclusion}}

We have identified a class of formal hamiltonians $\hat {\cal H}$ of one-dimensional
continuous quantum systems that feature energy point spectra which can be dialled up at
will.

Owing to the occurrence of high orders in momenta, the eigenfunctions of $\hat {\cal H}$
for these point spectra can be out of order with respect to the number of nodes associated
with level numbers (violation of Sturm-Liouville monotonic energy-level ordering rule).

We can however restrict the formal hamiltonians~$\hat {\cal H}_{(N)}$ to even or odd
expansion order~$N$ to enforce monotonic level ordering.
  
Our observations raise the question whether the construction of formal hamiltonians
$\hat {\cal H}$ provides a useful tool to universally represent and treat `all discrete
spectrum quantum systems' on an equal footing.

Investigation of the generalization of this approach to interacting multiparticle systems
appears warranted.

\begin{acknowledgments} O.~S. thanks Eran Ginossar for his suggestion to investigate Kerr
  systems. This work is partially supported by the Grant 254127 of CONACyT (Mexico).

 \end{acknowledgments}

\bibliography{Ole_Bibliography}

\end{document}